\documentclass[9pt]{extarticle}

\usepackage{sci_emu}

\begin{document}

\begin{multicols}{3}
\premulticols = 5pt
\postmulticols = 5pt
\multicolsep = 0pt

\columnbreak
{
\hangindent=\columnwidth

\noindent \begin{minipage}[c]{2.05\columnwidth}
\topic{\textcolor{red}{\fontfamily{phv}\fontseries{b}\selectfont GALAXY EVOLUTION}}
\title{Isolated compact elliptical galaxies: Stellar systems that ran away}
\author
{Igor Chilingarian$^{1,2}$\footnotemark
\, and Ivan Zolotukhin$^{3,2}$}
\date{}
\maketitle 


\begin{sciabstract} 
Compact elliptical galaxies form a rare class of stellar system ($\sim$30
presently known) characterized by high stellar densities and small sizes and
often harboring metal-rich stars.  They were thought to form through tidal
stripping of massive progenitors until two isolated objects were discovered
where massive galaxies performing the stripping could not be identified.  By
mining astronomical survey data, we have now found 195 compact elliptical
galaxies in all types of environment.  They all share similar dynamical and
stellar population properties.  Dynamical analysis for non-isolated galaxies
demonstrates the feasibility of their ejection from host clusters and groups
by three-body encounters, which is in agreement with numerical simulations.  
Hence, isolated compact elliptical and quiescent dwarf galaxies are tidally
stripped systems that ran away from their hosts.
\end{sciabstract}
\end{minipage}
%
}

\vskip 5pt
\renewcommand{\LettrineFontHook}{\fosfamily\fontseries{cl}\selectfont}
\lettrine[lines=5]{G}{} alaxies are thought to form through the hierarchical merging of smaller
building blocks into larger systems \cite{WF91,CLBF00} and the history of
these interactions is imprinted in their observable properties.  Some
galaxies such as ultra-compact dwarfs \cite{Drinkwater+03} and compact
ellipticals (cEs) \cite{Price+09,Chilingarian+09,CB10,Norris+14}, show
evidence of strong tidal interactions with massive neighboring galaxies
\cite{HPPH11} that stripped most of the stars from the compact galaxies'
progenitors. cEs are rare galaxies with high stellar densities that 
resemble centers of giant ellipticals but have masses that are about two 
orders of magnitude smaller [$M \sim 10^9$ solar mass ($M_{\odot}$)].  
They are found mostly in the cores of galaxy clusters next to massive 
central galaxies, which is in alignment with the above hypothesis for 
their evolution.

The recent discoveries of isolated cE galaxies \cite{HPP13,PLHH14} that do
not belong to any galaxy cluster or group raised another round of debate
about cE formation: whether they all formed through the tidal stripping
or through a different mechanism of formation such as mergers of dwarf
galaxies with specific morphologies and configurations \cite{PLHH14}. 
Dwarf--dwarf galaxy mergers do happen in vicinities of massive galaixes
\cite{Rich+12,AEvdV14}.  However, neither have they been observed in 
low--density environments, nor any of the remnants resemble properties of cE
galaxies.  The existence of a substantial number of isolated cEs will hence
imply notably higher dwarf--dwarf merger rates than predicted by
numerical simulations \cite{FMB10} and challenge the currently accepted
hierarchical structure formation paradigm.

We demonstrated that all known cE galaxies are outliers from the universal
optical--ultraviolet color--color--magnitude relation of 
\noindent \begin{minipage}[l]{\columnwidth}
{\vskip 5pt \begin{flushleft} \hrule {\fosfamily\fontseries{cl}\fontsize{7pt}{1em}\selectfont $^1$Smithsonian Astrophysical Observatory,
60 Garden Street MS09, Cambridge, MA 02138, USA.$^2$Sternberg Astronomical Institute, Moscow State University,
13 Universitetsky prospect, Moscow, 119992, Russia.$^3$L'Institut de Recherche en Astrophysique et
Plan\'etologie, 9, Avenue du Colonel Roche BP 44346, 31028, Toulouse Cedex 4,
France.\\{\fontseries{sb}\fontsize{5pt}{1em}\selectfont$^\ast$Corresponding author. E-mail:
igor.chilingarian@cfa.harvard.edu}} \end{flushleft}}
\end{minipage}

\columnbreak

\begin{minipage}[c][19.1\baselineskip]{\columnwidth}
\end{minipage}

\marginpar[aa]{bb}
\noindent galaxies \cite{CZ12}.  We could therefore perform a search for cE
galaxies not only in the centers of rich clusters and groups as has been
done before \cite{Chilingarian+09}, but across all environments using  data
from wide-field imaging surveys, the optical ground based Sloan Digital Sky
Survey [SDSS, \cite{SDSS_DR7}] and the ultra-violet all-sky survey carried
out by the GALaxy Evolution eXplorer [GALEX, \cite{Martin+05}] spacecraft,
which are all publicly available in the Virtual Observatory.

First, we created an initial list of candidates (supplementary materials)
from the sample of galaxies having spectra in the SDSS and, hence, known
distances, by selecting outliers above $+$0.035~mag in the optical $(g-r)$
color from the universal relation \cite{CZ12}.  We chose low luminosity
galaxies [$L < 4\times10^9$ solar luminosity $(L_{\odot})$ or absolute
magnitude $(M_g) > -18.7$~mag] that had small
half-light radii ($R_e < 0.6$~kpc) or were spatially unresolved in SDSS
images; did not show substantial ellipticity, which was essential for
removing edge-on spiral galaxies; had the redshifts in the range of $0.007 < z
< 0.08$ (distances between 30 and 340~Mpc); and either possessed red
near-ultraviolet colors [$(NUV - r) > 4$~mag] or remained undetected in the
$NUV$ band.  We constrained by color and also removed objects that have
emission lines in their spectra in order to exclude any objects with recent
or ongoing star formation.

We then fitted their SDSS spectra against a grid of stellar population
models using the ``NBursts'' code\cite{CPSK07} and obtained mean ages,
metallicities, and velocity dispersions of their stars.  We rejected
candidates with stellar ages younger than 4 billion years and introduced an additional
constraint based on stellar velocity dispersions ($\sigma > 60$~km/s). 
Stellar systems in equilibrium that are dynamically supported by random
motions of stars, as most elliptical galaxies are have their dynamical
masses ($M_{\rm{vir}}$), half-mass radii ($R_e$), and global velocity
dispersions ($\sigma_v$) connected by the simple virial relation:
$M_{\rm{vir}} = 9.96 R_e \sigma_v^2/G$ \cite{Spitzer69,Hernquist90}.  
Therefore, for a galaxy with known velocity dispersion and a stellar mass
($M_{*}$) derived from its luminosity and stellar population parameters, we
can estimate the lower limit for the half-light radius (if a galaxy contains
dark matter, its real half-light radius will be larger because
$M_{\rm{vir}}>M_{*}$).  Hence, we can firmly reject physically extended
objects such as ``normal'' dwarf elliptical galaxies that are unresolved in
SDSS images because of their large distances by selecting only objects with
high velocity dispersions.

In this fashion, we constructed a sample of 195 galaxies (Fig.~1 and
supplementary materials).  We then cross-matched this list against the SDSS
Galaxy Groups catalog \cite{TTL12} and established their group/cluster
membership.  For seven objects without counterparts in the Galaxy Groups
catalog, we identified possible host galaxies (in most cases, group centers)
located between 750~kpc and 3.3~Mpc in projection.  Because some bright and
extended galaxies were missing from the SDSS spectroscopic sample, and
therefore also from the Galaxy Groups catalogue, we used NASA/Infrared
Processing and Analysis Center (IPAC) Extragalactic Database (NED) for the
identification of host galaxies for 45 cEs.  Our final sample contains 56
galaxies in clusters, 128 in groups, and 11 isolated or field cE galaxies. 
Eight galaxies (supplementary materials) exhibit prominent tidal streams
similar to the two known cEs \cite{HPPH11}.

Ages and metallicities (Fig.~2 and supplementary materials), colors and luminosities of 11
isolated cE galaxies do not show a statistically significant difference from
those of galaxies being members of clusters and groups.  The
Kolmogorov-Smirnov (KS) probabilities of these properties for isolated and
non-isolated subsamples to be derived from the same underlying distribution
range from 30 (for luminosities) to 80\%\ (metallicities).  Our cE sample does
not show any correlation between the metallicity and the stellar mass
conversely to normal elliptical galaxies, which exhibit a rather tight
mass--metallicity relation (Fig.~2).

For rich groups and clusters hosting cE galaxies, we built caustic
diagrams\cite{Kaiser87,DG97,Diaferio99} that present differences of radial
velocities of cluster members from the cluster center versus projected
distances.  A galaxy position on such a diagram reflects its dynamical
status: objects sitting deep inside the cluster potential well are located
inside the distribution whereas galaxies near the edges are barely
gravitationally bound to the host cluster or infalling on to it for
the first time.

\begin{SCfigure*}
\includegraphics[width=2.48\columnwidth]{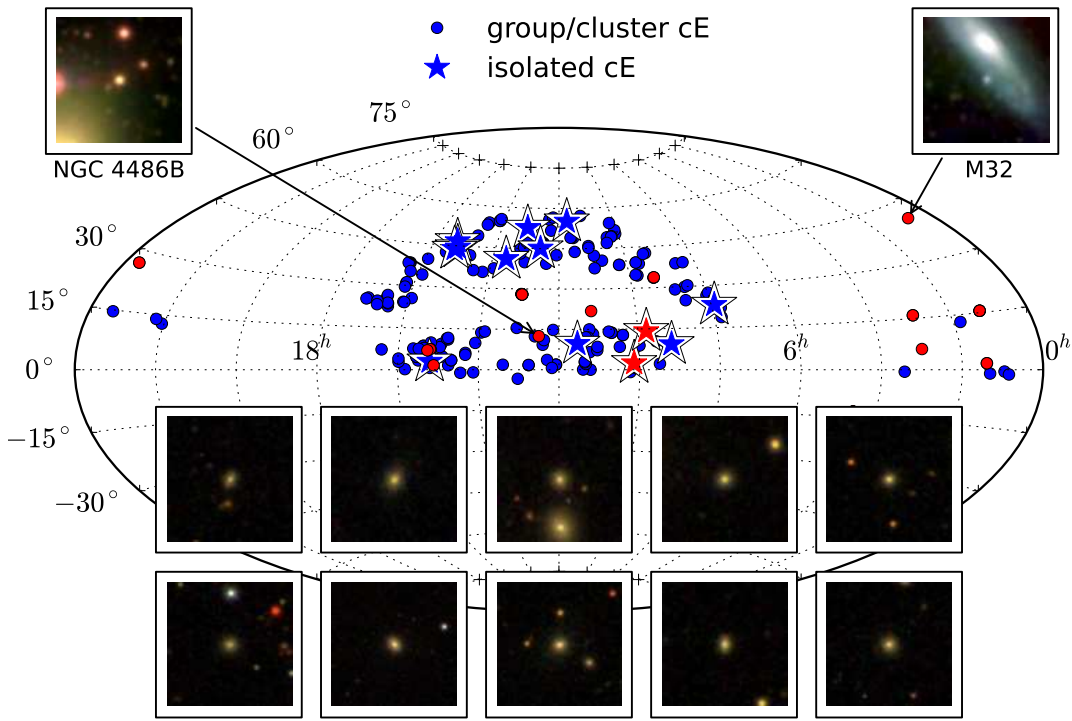}
\caption{\fontsize{8pt}{1.21em}\fontfamily{phv}\selectfont
{\fontseries{b}\selectfont A sample of compact elliptical galaxies in all types of
environment.} Our current sample of compact elliptical galaxies (blue
symbols) is compared with a dataset compiled from the literature (red symbols).
Dots and stars denote group/cluster and isolated compact elliptical
galaxies respectively. Square panels in the bottom part of the figure show
representatives of the current sample, and top corner insets display
Messier~32 and NGC~4486, prototypical cEs in the local Universe as they
would look with the SDSS telescope at a 130~Mpc distance ($z=0.03$). Each inset 
panel covers a 20 by 20~kpc region centered on a cE.}
\end{SCfigure*}

We constructed an ensemble cluster by normalizing individual cluster and
group data by their velocity dispersions and sizes for 33 structures from
our sample each of which included over 20 member galaxies \cite{CYE97,BG03}. 
Then we computed its caustic diagram in order to visualize the phase space
pattern of the infalling galaxy population and overplotted our cE galaxies
on it (Fig.~3).  The cE population strongly differs from other cluster
members.  The KS tests for projected distance and radial velocity
distributions reject the hypothesis of cE and cluster member samples being
derived from the same parent population at the 97 and 98\%\ levels. 
Numerical simulations of tidal stripping\cite{Chilingarian+09,PB13} suggest
that a progenitor galaxy, even if it approaches a cluster center on a very
extended radial orbit, will lose a major fraction of its orbital energy
because of dynamical friction, become gravitationally locked in the inner
region of a cluster on a tightly bound orbit, and will finally be accreted
by the host galaxy after a few billion years.  Many cEs from our sample
exhibit this behaviour (Fig.~3 and supplementary materials).  However, we
see a number of cE galaxies close to the edges of the caustics suggesting
that they are barely gravitationally bound to the cluster potential because
they do not belong to the infalling population as we demonstrated.  This
looks completely unrealistic in the case of a one-to-one galaxy encounter
resulting in tidal stripping, but in the case of a three- or multiple-body
encounter this situation becomes significantly more likely.

An interaction of binary stars with the cen-
\columnbreak
{
\noindent \begin{minipage}[t][26.9\baselineskip]{2.05\columnwidth}
\vskip -22pt
\floatsetup[figure]{floatwidth=.73\columnwidth,capposition=beside,capbesideposition={top,left}}
\begin{figure}[H] 
\vskip -20pt
\includegraphics[width=1.025\hsize]{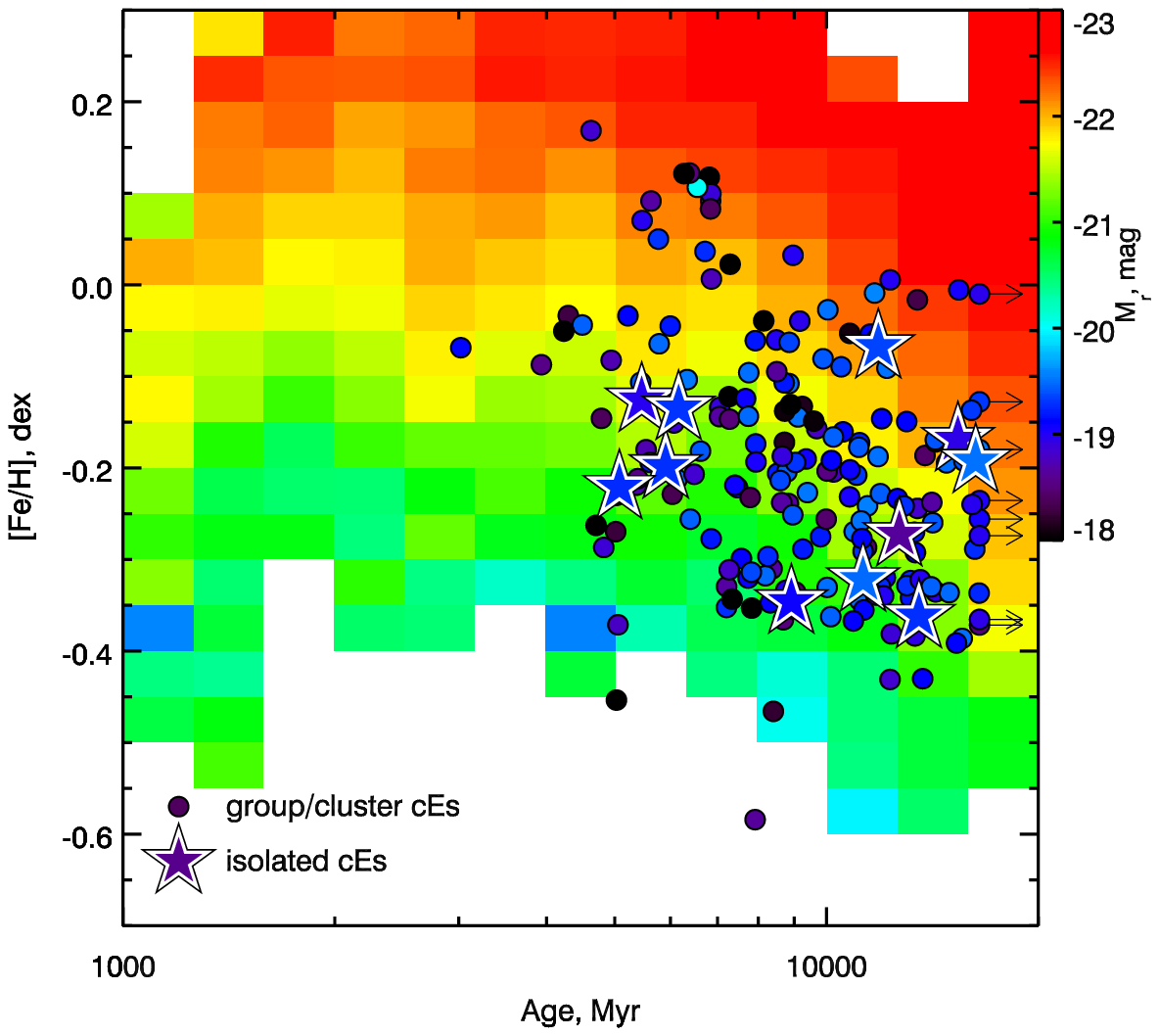}
\caption{\fontsize{8pt}{1.21em}\fontfamily{phv}\selectfont
{\fontseries{b}\selectfont  Comparison of stellar population properties of
isolated and nonisolated compact elliptical galaxies to a reference
sample of elliptical galaxies from the SDSS}. Mean
stellar ages and metallicities of cE galaxies hosted
by groups and clusters (round symbols) and isolated cEs (stars) were
determined from the modeling of their SDSS spectra by using the
``NBursts'' full spectrum 
fitting. The plotting symbols}
\end{figure}

{\fontsize{8pt}{1.15em}\fontfamily{phv}\selectfont
\begin{flushleft}
\vskip -1.9em
\noindent  
are color coded 
according to the $r$-band absolute magnitudes derived from
the SDSS photometry. We also computed ages and metallicities for a reference
sample of 82500 elliptical galaxies from SDSS DR7 in the same fashion along
with the median $r$-band absolute magnitude for every bin of age--metallicity
parameter space. These magnitudes are shown as a background map, with the
colors on the same scale as cE galaxies. This figure demonstrates
that: (i) stellar populations of isolated and group/cluster cEs do not
differ statistically and (ii) cEs are on average much fainter than are normal
elliptical galaxies of the same stellar age/metallicity.
\end{flushleft}
}
\end{minipage}
}

\noindent tral supermassive black hole is
one accepted scenario for the creation of hypervelocity stars \cite{BGKK05}
in our Galaxy: One of the binary components is ejected, whereas the other one
falls on-to the black hole.  Numerical simulations 

\columnbreak
{
\begin{minipage}[t][27.3\baselineskip]{1.0\columnwidth}
\end{minipage}
}

\noindent suggest\cite{SNAS07} that
three-body encounters are responsible for putting Milky Way satellites on
extreme orbits going as far as 3~Mpc away.  Even though typical galaxy
clusters have much wider and deeper potential wells than that of the Local Group,
three- and even multiple-body encounters must happen much more frequently in
those dense environments.  Therefore a certain probability exists that some
of them will lead to the gravitational ejection of galaxies participating in
the interaction to extreme radial orbits with the apocentric distances of a
few megaparsecs \cite{WTCB14}.  A three-body encounter that might eject a cE galaxy
from its host cluster or group does not have to happen during the cE
formation through tidal stripping, that is two galaxies do not have to fall on-to
the cluster/group center at the same time.  When a cE progenitor is tidally
stripped, it quickly settles on a tightly-bound rapidly decaying orbit
\cite{Chilingarian+09} and if another galaxy infalls later, but before a
newly formed cE has been accreted (hundreds of millions to a couple of
billion years), the three-body encounter becomes possible. 

We estimate the probability of a close three-body encounter geometrically. 
Numerical simulations suggest \cite{DLB07} that over a typical cE lifetime
of 2 billion years \cite{Chilingarian+09}, an average brightest cluster
galaxy (BCG) must have experienced three of four mergers with massive ($M \gtrsim
10^{10}~M_{\odot}$) galaxies.  We assume that: (i) a typical cE resides on a
quasicircular orbit within $r_{cE}\sim$120~kpc from a host BCG galaxy after
correction for projection effects (fig.~S4), (ii) galaxies infall on a
BCG on radial orbits from random directions, and (iii) a three-body encounter will
be sufficiently close if a cE passes within $r_{3b}\approx20$~kpc from a
massive infalling galaxy.  Hence, the probability is as a volume ratio of a
cylinder of radius $r_{3b}$, height $r_{cE}$ and a sphere of radius
$r_{cE}$, $P_{3b} = 3/4 (r_{3b}/r_{cE})^2 \approx 0.02$, or $\sim$6 to 8\%\
for three of four merger events.

In our sample of cluster and group cE galaxies, we indeed see numerous
examples in which a cE resides only 20 to 80~kpc in projection from an ongoing
major merger scene or several other massive cluster members are visible in
the cE vicinity apart from the massive central cluster/group galaxy.  Also,
there is a known example of a globular cluster in the Virgo cluster
\cite{Caldwell+14} that was likely ejected at the speed of 2500~km/s and
became gravitationally unbound to the cluster and its central galaxy
Messier~87.

We conclude that the tidal stripping process can explain all observational
manifestations of compact elliptical galaxies, including the formation of
isolated cEs whose existence was suggested as a strong counter-argument for
tidal stripping \cite{HPP13}.  The ejection of cEs from central regions of
galaxy clusters by three-body encounters is a channel for these galaxies to
survive for an extended period of time in the violent cluster environment
where they would otherwise be accreted by massive hosts on a timescale of 2
billion to 3 billion years.  
\columnbreak
{
\noindent \begin{minipage}[t][43.9\baselineskip]{2.05\columnwidth}
\vskip -19pt
\floatsetup[figure]{floatwidth=.73\columnwidth,capposition=beside,capbesideposition={top,left}}
\begin{figure}[H] 
\vskip -20pt
\makebox[\hsize]{\hspace*{-10pt}\includegraphics[width=1.05\hsize]{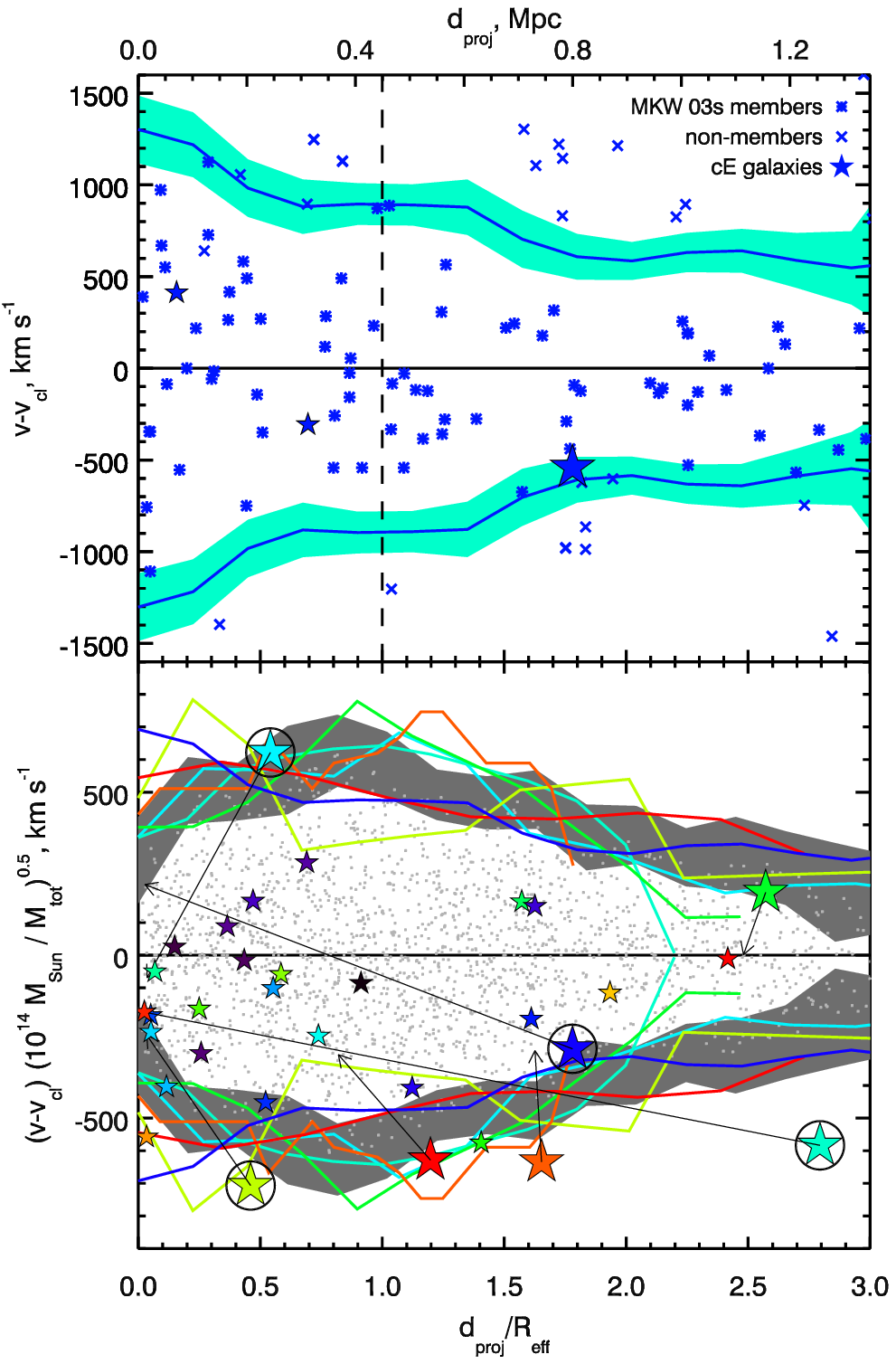}}
\caption{\fontsize{8pt}{1.21em}\fontfamily{phv}\selectfont
{\fontseries{b}\selectfont  
Positions of cE galaxies identified on caustic diagrams in
galaxy clusters and rich groups with more than 20 members in the SDSS.}
({\fontseries{b}\selectfont Top}) An example of a caustic diagram showing projected distances and
radial velocity differences of the members of the cluster MKW~03s that
includes cE galaxies shown as stars.  The derived caustics shown as solid
blue lines roughly correspond to the escape velocity for a galaxy at a given
distance from the cluster center.  Shaded areas show statistical
$1~\sigma$ uncertainties of the caustic line computation.
({\fontseries{b}\selectfont Bottom}) The
caustic lines for an ensemble cluster of 2,592 individual members are shown
as gray shaded areas and light gray dots, respectively.  The caustic lines
normalized by}
\end{figure}

{\fontsize{8pt}{1.15em}\fontfamily{phv}\selectfont
\vskip -16pt
\begin{flushleft}
\vskip -2.9em
\noindent corresponding velocity dispersions  and half-mass radii
($R_{\mbox{eff}}$) are shown for six individual clusters as colored solid
lines.  Small stars denote cEs located deeply inside the potential wells of
these six clusters, and large stars (with same colors as caustic lines) 
indicate cEs that have projected radial velocities of at least 85\% of the caustic
amplitude.  Vectors point to the location of identified host galaxies for
corresponding cEs.  Three of them are different than the cluster central
galaxies indicating that these cEs belong to subgroups inside clusters and
that the caustic diagnostic is irrelevant for them. However, the four circled
cEs are examples of galaxies barely gravitationally bound to their host
clusters.
\end{flushleft}
}
\end{minipage}
}

\vskip 12pt
\noindent The 11 isolated cEs probably represent a
population of runaway galaxies that received sufficient kick velocities to
leave their host clusters or groups forever.

The gravitational ejection mechanism may also explain the very existance of
extremely 

rare isolated quiescent dwarf galaxies \cite{GBYT12}, where the
star formation quenching is usually explained by environmental effects. 
These systems are more spatially extended than cEs and do not exhibit
substantial tidal stripping footprints.  This suggests that they never came
very close to cluster/group centers and therefore, the three-body
encounter probability for them should be lower than that for cEs although
still nonnegligible.

\columnbreak
\begin{minipage}[t][44.2\baselineskip]{\columnwidth}
\end{minipage}

\bibliography{cE_GALEXSDSS}
\bibliographystyle{Science_mod}


\section*{ACKNOWLEDGMENTS}
{\fosfamily\fontseries{cl}\fontsize{7pt}{7pt}\selectfont 
The authors are greatful to F.~Combes (Observatoire de Paris), I.~Katkov
(Sternberg Astronomical Institute), and M.~Kurtz (Smithsonian Astrophysical
Observatory) for useful discussions and critical reading of the manuscript. 
This result emerged from the tutorial run by the authors at the Astronomical
Data Analysis Software and Systems conference in 2012.
The authors acknowledge support by the Russian Science Foundation project
14-22-00041 ``VOLGA -- A View On the Life of GAlaxies'' .  The project used
computational resourses funded by the M.V.Lomonosov Moscow State University
Program of Development.  This research has made use of Aladin developed by
the Centre de Donn\'ees Astronomiques de Strasbourg; TOPCAT and STILTS
software packages developed by M.Taylor; ``exploresdss'' script
by G.~Mamon; the VizieR catalogue access tool, CDS, Strasbourg, France; and
the NASA/IPAC NED which is operated by the Jet
Propulsion Laboratory, California Institute of Technology, under contract
with the NASA.  Funding for the
SDSS and SDSS-II has been provided by the Alfred P.  Sloan Foundation, the
Participating Institutions, the National Science Foundation, the U.S. 
Department of Energy, NASA, the
Japanese Monbukagakusho, the Max Planck Society, and the Higher Education
Funding Council for England.  The SDSS Web Site is
\url{http://www.sdss.org/}. 
GALEX and SDSS databases used in our study are available via the CasJobs
web-site \url{http://skyserver.sdss.org/CasJobs}.}

\section*{SUPPLEMENTARY MATERIALS}
{\fosfamily\fontseries{cl}\fontsize{7pt}{7pt}\selectfont 
\url{http://www.sciencemag.org/content/348/6233/418/suppl/DC1}\\
Materials and Methods\\
Supplementary Text\\
Figs. S1 to S4\\
Table S1\\
\vskip 5pt \noindent 19 November 2014; accepted 20 March 2015\\
10.1126/science.aaa3344
}



\end{multicols}

\clearpage

\section*{SUPPLEMENTARY MATERIALS}
\renewcommand\thesubsection{S\arabic{subsection}}
\renewcommand\thefigure{S\arabic{figure}}
\renewcommand\thetable{S\arabic{table}}
\setcounter{figure}{0}

\begin{multicols}{3}
\subsection{Input galaxy catalogue}

As a starting point for the search of cE galaxy candidates, we compile
the input galaxy catalogue. In order to do this, we use publicly available datasets,
GALEX Data Release 6 \cite{Martin+05}, SDSS Data Release 7 \cite{SDSS_DR7}
and UKIDSS Data Release 8 \citesm{Lawrence+07}.  Initially, from the SDSS DR7
spectral sample we select normal non-active galaxies with the redshift in
the range $0.007 < z < 0.6$.  This list is cross-matched with the UKIDSS Large
Area Survey catalogue using the best match criterion with the 3~arcsec radius. 
The SDSS--UKIDSS sample is then uploaded to the GALEX CasJobs service in order
to look for best matches within 3~arcsec.  As a result of this procedure, we
obtain a catalogue of 429\,707 galaxies with fully corrected
UV-to-optical spectral energy distributions in 11 photometrical bands ($FUV$
and $NUV$ from GALEX, $ugriz$ from SDSS, and $YJHK$ from UKIDSS) and
spectra.  Petrosian magnitudes are $k$-corrected using 
analytical approximations \citesm{CMZ10}.  SDSS spectra are fitted
with state-of-the-art stellar population models. This procedure
yields ages and metallicities of their stars as well as stellar
velocity dispersions which we add to the catalogue.  

\subsection{Full spectral fitting}

\begin{figure}
\includegraphics[width=0.9\hsize]{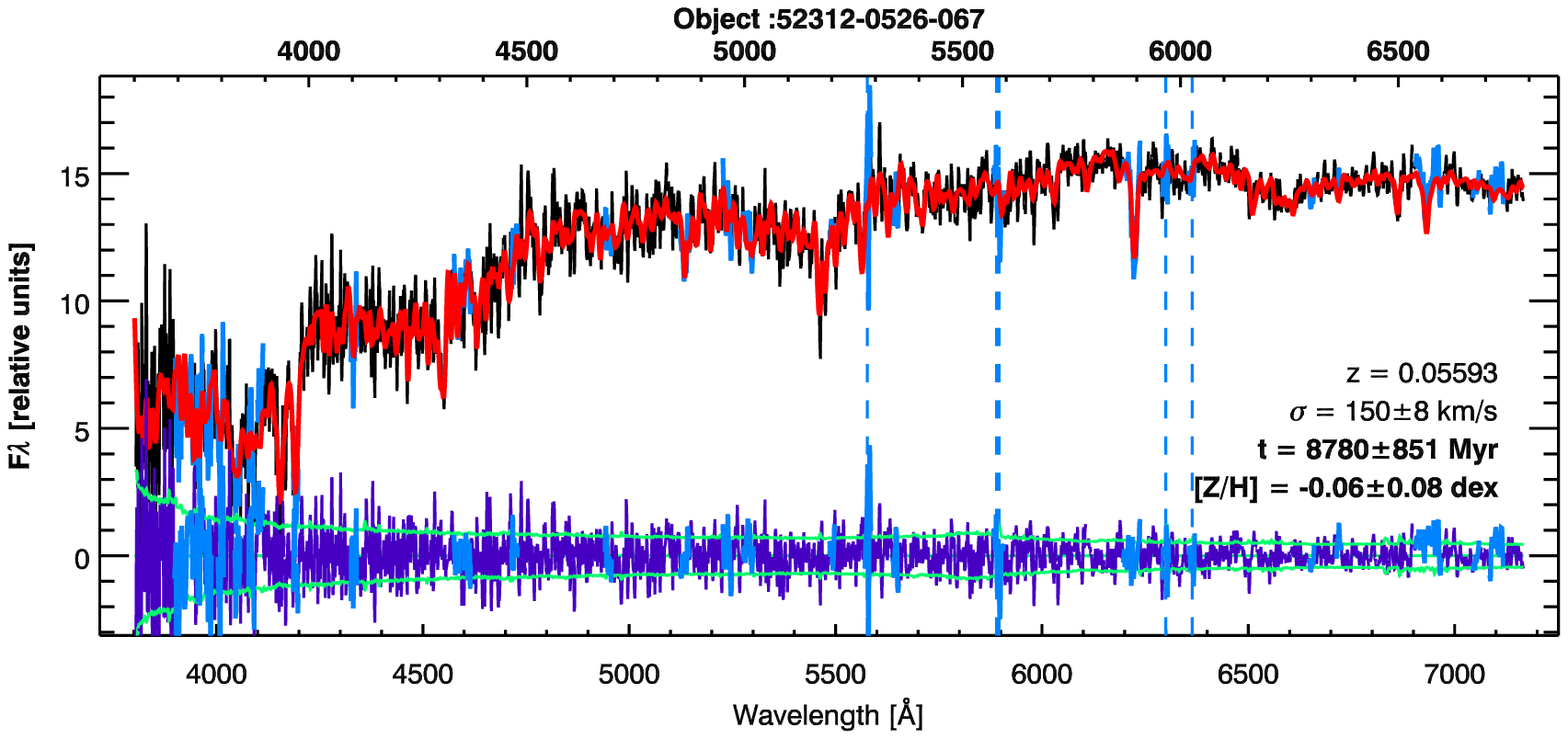}
\caption{\fontsize{8pt}{1.21em}\fontfamily{phv}\selectfont An example of the full spectral fitting of an SDSS DR7 spectrum of
a compact elliptical galaxy with the ``NBursts'' technique. The black,
green, red and purple curves represent the observed spectrum, its
uncertainties ($\pm 1 \sigma$), the best-fitting model and
residuals. Position of prominent atmosphere airglow lines are indicated by
dashed blue lines.\label{spec_example}}
\end{figure}

We use the ``nbursts'' full spectral fitting technique \cite{CPSK07} with
high-resolution ($R=10,000$) {\sc pegase.hr} \citesm{LeBorgne+04} simple
stellar population (SSP) models computed using the Salpeter stellar initial
mass function \citesm{Salpeter55} in order to extract kinematics and stellar
population properties from our spectra (see Fig.~\ref{spec_example}).

The fitting procedure comprises the following steps: (1) a grid of SSP
spectra with a fixed set of ages (spaced logarithmically from 20~Myr to
18~Gyr) and metallicities (from $-$2.0 to $+$0.5~dex) is convolved with the
wavelength-dependent instrumental response of SDSS provided along with every
spectrum in the 6th {\sc fits} extension; (2) a non-linear least square
fitting against an observed spectrum is done for a template picked from the
pre-convolved SSP grid interpolating on age ($\log t$), and metallicity
($Z$), broadened with the line-of-sight velocity distribution (LOSVD)
parametrized by $v$ and $\sigma$ and multiplied pixel-by-pixel by an
$n^{\rm{th}}$ order Legendre polynomial (multiplicative continuum),
resulting in $n + 5$ parameters to be determined (we used $n=19$ for SDSS
spectra). 

The fitting procedure, the error analysis and degeneracies between
kinematics and stellar population parameters have been throughly described
\citesm{CPSA07}.  We analysed the systematics of stellar population
parameters with respect to the $\alpha$-element enhancement
\citesm{Chilingarian+08}, the sensitivity of different spectral features
\citesm{Chilingarian09}.  The choice of the stellar initial mass function in
the models is shown not to affect stellar metallicity estimates and only
slightly affect stellar ages \citesm{PCK13}.

In order to precisely estimate velocity dispersion, we convolve the SSP grid
with the spectral line spread function (LSF) of the spectrograph varying
across the wavelength range.  In SDSS DR9 spectra, the LSF shape is
available in the 6th FITS extension of every data file.  It was demonstrated
\citesm{CMHI11} that for intermediate signal-to-noise spectra the stellar
velocity dispersion can be estimated with 10\%\ uncertainties down to
$\sim$1/2 of the LSF width (expressed as Gaussian $\sigma$). For SDSS this
corresponds to $\sim45$~km/s that is substantially below our velocity
dispersion selection criterion ($\sigma>60$~km/s, see below).

\subsection{Selection of candidate compact elliptical galaxies}

By placing known cE galaxies on the 3D optical-UV color--color--magnitude
diagram we noticed that they occupy a distinct region in this parameter
space \cite{CZ12}.  We, hence, use several approaches to define this region
in order to make it suitable for a formal database query.  They result in 5
samples of cE candidates extracted from the catalogue described above. 
Since the input data were generated by several sometimes imperfect data reduction
pipelines, we have to visually inspect every candidate image in order to
discard extended galaxies with bright nuclei, edge-on spiral galaxies and
artefacts, which also satisfied our initial automated cE selection criteria.

We selected objects that have: (a) red optical $(g-r)$ colors, at least $+0.035$~mag
above the universal color--color--magnitude relation \cite{CZ12}; (b)
low luminosities [$L < 4\cdot10^9 L_{\odot}$ or $M_g > -18.7$~mag]; (c)
small half-light radii [$R_e < 0.6$~kpc] or remain spatially unresolved in SDSS
images; (d) no significant ellipticity; (e) the redshifts in 
the range $0.007 < z < 0.08$ [distances between 30 and 340~Megaparsecs]; (f)
either possess red near-ultraviolet colors [$(NUV - r) > 4$~mag] or
drop out of the $NUV$ band; (g) no significant emission lines in
their spectra; (h) velocity dispersions $\sigma>60$~km/s and stellar
population ages $t>4$~Gyr.

In the next subsections we list the criteria that we used to build those
samples.

\end{multicols}

\subsubsection{Initial sample}

The initial sample included 108 galaxies that have $NUV-r > 4.0$~mag, the
absolute $g$ magnitude fainter than -18.7~mag, the 50~per~cent Petrosian
radius of smaller than 2~arcsec in the angular measure or below 0.7~kpc in the
physical scale, a distance from the 3D surface \cite{CZ12} along the $g-r$
color axis greater than 0.035~mag, the stellar age older than 4~Gyr and the
velocity dispersion above 60.0 km~s$^{-1}$, the latter two parameters
derived from the SSP fitting.

Below we give the formal specification of this filter using subsets
expressions from the {\sc topcat} table processing software \citesm{taylor05}
with the column names from the catalogue.  Note that {\sc gr\_fit} column is
a 3D surface value in ($g-r$, $NUV-r$, $M_g$) coordinates.

\small
\begin{verbatim}
(corrmag_NUV - corrmag_r - kcorr_NUV + kcorr_r) > 4.0 &&
(corrmag_g - kcorr_g - 25 - 5 * log10(luminosityDistance(z, 72.0, 0.3, 0.7))) > -18.7 && 
(petror50_r < 2.0 || petror50_r / 206.265 * luminosityDistance(z, 72.0, 0.3, 0.7) < 0.7) && 
(corrmag_g - corrmag_r - kcorr_g + kcorr_r - gr_fit) > 0.03 &&
age_ssp > 4000.0 && dispvel_ssp > 60.0
\end{verbatim}
\normalsize

After visual inspection, we have rejected 52 contaminated objects and
therefore 56 cE galaxies were left in this sample.  The latter two criteria
on the age and velocity dispersion ({\tt age\_ssp > 4000.0 \&\& dispvel\_ssp
> 60.0}) are common between all samples except the sample selected by the virial
radius criterion.

\subsubsection{UV-dropout sample}

The second sample is constructed with a relaxed requirement of a candidate
detection in the $NUV$ band, because only 47~per~cent of galaxies from our
catalogue have $NUV$ magnitudes.  We dropped the constraints on the $NUV-r$ color
and the one derived from the 3D photometric relation (because it also requires
a $NUV$ magnitude).  Instead, together with our usual requirement of a galaxy
having SSP age older than 4~Gyr and the velocity dispersion higher than
60~km~s$^{-1}$, we added a simple geometrical constraint from the canonical
color--magnitude $(g-r, M_r)$ diagram: a candidate galaxy is redder than
the red sequence plus $1\sigma$ scatter.  This translates into the following
expression: $g-r > 0.8 - (M_r+20.0) \cdot 0.1 / 3.0$.  Hence, we obtain the
following formal query expression:

\small
\begin{verbatim}
(corrmag_g - kcorr_g - 25 - 5 * log10(luminosityDistance(z, 72.0, 0.3, 0.7))) > -18.7 &&  
(petror50_r < 2.0 || petror50_r / 206.265 * luminosityDistance(z, 72.0, 0.3, 0.7) < 0.7) &&  
age_ssp > 4000.0 && (dispvel_ssp > 60.0 || dispvel_exp > 60.0) && 
(corrmag_NUV < -100 || NULL_corrmag_NUV) && 
(corrmag_g - corrmag_r - kcorr_g + kcorr_r > 0.8 - 
(corrmag_r - kcorr_r - 25 - 5 * log10(luminosityDistance(z, 72.0, 0.3, 0.7)) + 20.0) * 0.12 / 3.0)
\end{verbatim}
\normalsize

This query yields 288 objects. After visual inspection we added 32 additional cE candidates to our initial sample.

\subsubsection{Estimated effective radius sample}

We notice that well known cE galaxies Messier~32, NGC~4486B, NGC~5846cE have
effective radii smaller than 0.4~kpc.  We therefore try to visually inspect
galaxies that satisfy our initial selection criteria but instead of
filtering the input catalogue by the velocity dispersion estimate, we
require galaxies to have the effective radius estimate below 0.5~kpc,
similarly to known cEs.  In order to do this, we compute mass-to-light
ratios for galaxies from the input catalogue using the {\sc pegase.hr}
stellar population models constructed using the Salpeter IMF
\citesm{Salpeter55} and then estimate effective radii using the published
expression \cite{Spitzer69} assuming that mass follows light.  This filter
can be formally expressed as

\small
\begin{verbatim}
(corrmag_NUV - corrmag_r - kcorr_NUV + kcorr_r) > 4.0 && 
(corrmag_g - kcorr_g - 25 - 5 * log10(luminosityDistance(z, 72.0, 0.3, 0.7))) > -18.7 && 
(petror50_r < 2.0 || petror50_r / 206.265 * luminosityDistance(z, 72.0, 0.3, 0.7) < 0.7) && 
(corrmag_g - corrmag_r - kcorr_g + kcorr_r - gr_fit) > 0.03 && 
age_ssp > 4000.0 && r_eff < 0.5 && z < 0.08
\end{verbatim}
\normalsize

This query returned 82 objects from which we visually confirmed two.

To relax the requirement on having a NUV magnitude, we change the virual
radius query in a way we did for the UV-dropout sample.  The corresponding
estimated effective radius selection hence becomes:

\small
\begin{verbatim}
(corrmag_g - kcorr_g - 25 - 5 * log10(luminosityDistance(z, 72.0, 0.3, 0.7))) -18.7 &&  
(petror50_r < 2.0 || petror50_r / 206.265 * luminosityDistance(z, 72.0, 0.3, 0.7) < 0.7) &&  
age_ssp > 4000.0 && (corrmag_NUV < -100 || NULL_corrmag_NUV) && 
(corrmag_g - corrmag_r - kcorr_g + kcorr_r > 0.8 - 
(corrmag_r - kcorr_r - 25 - 5 * log10(luminosityDistance(z, 72.0, 0.3, 0.7)) + 20.0) * 0.12 / 3.0)  && 
r_eff_virial < 0.5 && z < 0.08
\end{verbatim}
\normalsize

This query returns 102 objects. Five of them are confirmed visually as cE
candidates and included in the preliminary list.

\vskip 10pt
\begin{multicols}{3}
We attempted to use the effective radii computed by the two-dimensional
fitting of SDSS images with the {\sc gim2d} software \citesm{Simard+11}.
However, in the vast majority of cases our galaxies remain spatially
unresolved in SDSS images, and the effective radii estimated from the image
analysis \citesm{Simard+11} do not correspond to their physical radii. This
is illustrated by the fact that they are correlated with redshift.
Therefore, we decided not to use these data in our further selection and
analysis.

\subsubsection{Additional candidates}

We plot the color--magnitude ($g-r, r$) diagram and visually inspect all
the objects with unphysically red colors (also satisfying common age and
velocity dispersion criteria), some of which turn out to be galaxies
blended with bright stars, artefacts and galaxies with underestimated $g$
magnitudes by 0.3~mag or more.  Among the latter group we find 120 cE
candidates which we include in the preliminary list of cE candidates.

\subsubsection{Discarding objects with emission lines}

After merging 4 lists of cE candidates selected using different approaches
described above, we obtain one preliminary table with 215 cE
candidates.  We then perform the final quality check by identifying
galaxies with emission lines in the spectra from this preliminary list of
candidates.  We compute line fluxes in the spectral fitting residuals and
use original flux uncertainties in order to estimate the signal-to-noise
ratios.  Then, we reject the objects where either [O{\sc ii}] 3730\AA, or
[O{\sc iii}] 5007\AA, or H$\alpha$ 6565\AA, or [N{\sc ii}] 6585\AA\, exceed
the 4-$\sigma$ detection (20 galaxies). 

Hence, we end up with 195 objects in our final sample of cE galaxies.

\subsection{Identification of host galaxies}

The reliable formal identification of host galaxies is a challenging
procedure and is not always possible.  We employ a manual and to some extent
subjective scenario and note that our host identification may be imperfect
in cases of objects with large projected distances from hosts.

In order to achieve this, we inspect SDSS color images in 2~degree
vicinities of every cE candidate using the {\sc cds aladin} software.  In
this procedure, we overplot all SDSS objects with known spectroscopic
redshifts within 2000~km~s$^{-1}$ of the radial velocity difference from a
cE galaxy.  Among those we visually identify the most luminous galaxy in a
group/cluster where it is possible.  In cases where the largest group/cluster member has no
SDSS redshift (e.g.  due to a fiber collision in the SDSS spectroscopic
targets selection procedure), we query NASA NED service to obtain its
redshift.  In a few cases, we select a close-by bright galaxy sometimes
showing signs of interaction with a cE candidate instead of the brightest
object in the group/cluster.  In our preliminary list of cE galaxies there
are 12 isolated or ``field'' objects for which we cannot identify hosts,
whereas in the discarded emission line galaxies there is 1 field galaxy. 
Hence, in the final sample we have 11 isolated cE galaxies.

\subsection{Properties of the final cE sample}

\subsubsection{Compact elliptical galaxies with tidal streams}

\begin{figure}
\includegraphics[width=0.9\hsize]{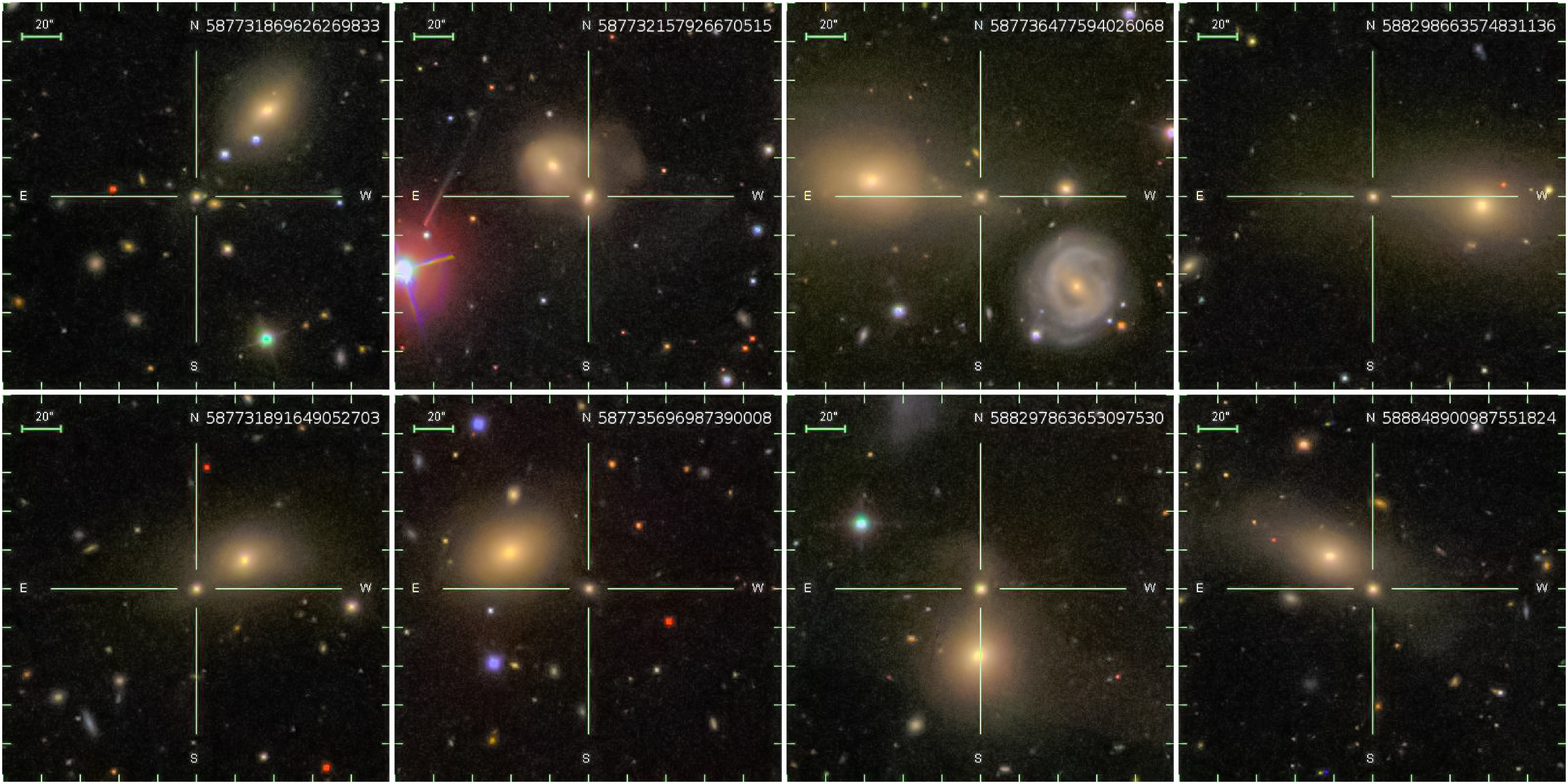}
\caption{\fontsize{8pt}{1.21em}\fontfamily{phv}\selectfont Compact elliptical galaxies showing signs of ongoing tidal
interactions with their hosts.\label{cEtidalfig}}
\end{figure}

Our final sample includes 8 compact elliptical galaxies with the visual
signs of ongoing tidal interaction with their host galaxies (see
Fig.~\ref{cEtidalfig}) resembling the two galaxies with tidal streams found
earlier \cite{HPPH11}.  The numerical simulations of the tidal stripping
\cite{Chilingarian+09} demonstrated that such features can be observed only
during a relatively short stage of the cE formation, that is during the
first 200--400~Myr following the first close passage of a cE progenitor near
the host galaxy.  Later they quickly fade down to very faint surface
brightness values ($>$29~mag~arcsec$^{-2}$) and become a part of the
intracluster (intragroup) light.

\subsubsection{Host galaxies}

Our new large cE sample allows us to compute statistical properties of the
host galaxies identified for 184 cEs.  Most cEs in our sample belong to
galaxy groups including 5 to 30 members identified in the SDSS spectroscopic
sample.  For 23 cEs, host galaxies are early to late-type spirals (in all
these cases there is only one cE associated to a given host galaxy) while
for the remaining cases the hosts are massive lenticular, elliptical or
merging galaxies.

A distribution of the differences between $r$-band luminosities of cE
galaxies and their hosts will reflect the distribution of stellar mass
ratios because stellar $r$-band $M/L$ ratios of cEs and giant early-type
galaxies are similar (assuming the same stellar initial mass function).  The
obtained $M_r(cE)-M_r(host)$ distribution is shown in
Fig.~\ref{figMMhost}.  It is skewed and peaks at 3.25~mag
corresponding to the stellar mass ratio of 20.  Giant elliptical galaxies
contain about 90\%\ of dark matter within their virial radii
\citesm{LW99,ML05b}, while for nearby cEs like \emph{Messier~32} the measured
dark matter content is negligible being consistent with the tidal stripping
scenario where the extended dark matter halo is stripped before the stellar
component.  This yields in the median cE-to-host mass ratio of about 200.

Worth mentioning that prototypical cEs, \emph{Messier~32} and
\emph{NGC~4486B} with the cE-to-host stellar mass ratios of 100 and 60
respectively are sitting in the tail of the distribution including a few
per~cent of objects.

In Fig.~\ref{figdproj} we show the distribution of projected
distances on the sky between cEs and their identified host galaxies.  The
distribution looks almost uniform between 20~kpc and 1~Mpc (in $\log
d_{\rm{proj}}$) with a prominent peak at 50--100~kpc.  The numerical
simulations of tidal stripping \cite{Chilingarian+09} demonstrate that if a
cE progenitor comes too close to the host galaxy, the quick orbital decay
causes its very quick accretion on-to the host on a timescale of a few
hundred Myr.  On the other hand, under typical conditions in clusters and
rich groups, tidal stripping becomes significantly less efficient at
distances exceeding 200~kpc, and this explains the decline. The
high-$d_{\rm{proj}}$ tail probably contains mostly cEs which were
gravitationally ejected from the cluster/group centers by three-body
interactions.

\subsection{Construction of caustics for cE host clusters and groups}

We identify 191 cE galaxies as members of galaxy clusters from the groups
and clusters catalogue of \cite{TTL12}.  7 of them have the luminosity rank
1 in a group of 2 galaxies and, therefore, were considered isolated or field
cEs.  In order to study the dynamical connection between cE galaxies and
their host environments, we plot phase space diagrams showing redshifts of
clusters members versus their projected distances from cluster centers. 
Such diagrams usually show a characteristic trumpet shape while their
boundaries are called caustics.  \cite{Diaferio99} provided a technique to
compute them under assumptions of spherical symmetry and hierarchical
clustering for the formation of the large-scale structure.  This method
works as a convenient way to estimate the escape velocity from the
gravitational potential well generated by the cluster at each
cluster-centric radius.  We used the implementation of \cite{Diaferio99}
technique in the Caustic App version 1.2 \citesm{CausticApp} to find caustics in the data
for our galaxy clusters and identify cE positions with
respect to the caustics.

We also create an ensemble galaxy cluster to compare the distribution of
normal cluster members to the one of our cE candidates in the phase space. 
We take 33 galaxy clusters that have more than 20 members and include at
least one cE candidate.  Their masses, velocity dispersions and effective
radii were already known after the previous step when their individual
caustics were computed and analysed.  We project every cluster on a
tangential plane with the projection center at the cluster center, and then
normalize projected coordinates of galaxies by half-mass radii of the
corresponding clusters.  The relative velocities with respect to a cluster
center are normalized by the cluster velocity dispersions.  Then, all the
clusters are shifted to the redshift $z_{ens}=0.05$.  In this fashion, positions and
radial velocities of 2,592 galaxies in 33 galaxy clusters are brought to
the same scale ready to be used as ensemble cluster.  We computed caustics
of generated ensemble cluster using default settings of Caustic App.
\end{multicols}


    \centering
    \begin{minipage}{.45\textwidth}
\begin{figure}[H]
        \centering
\includegraphics[width=\hsize]{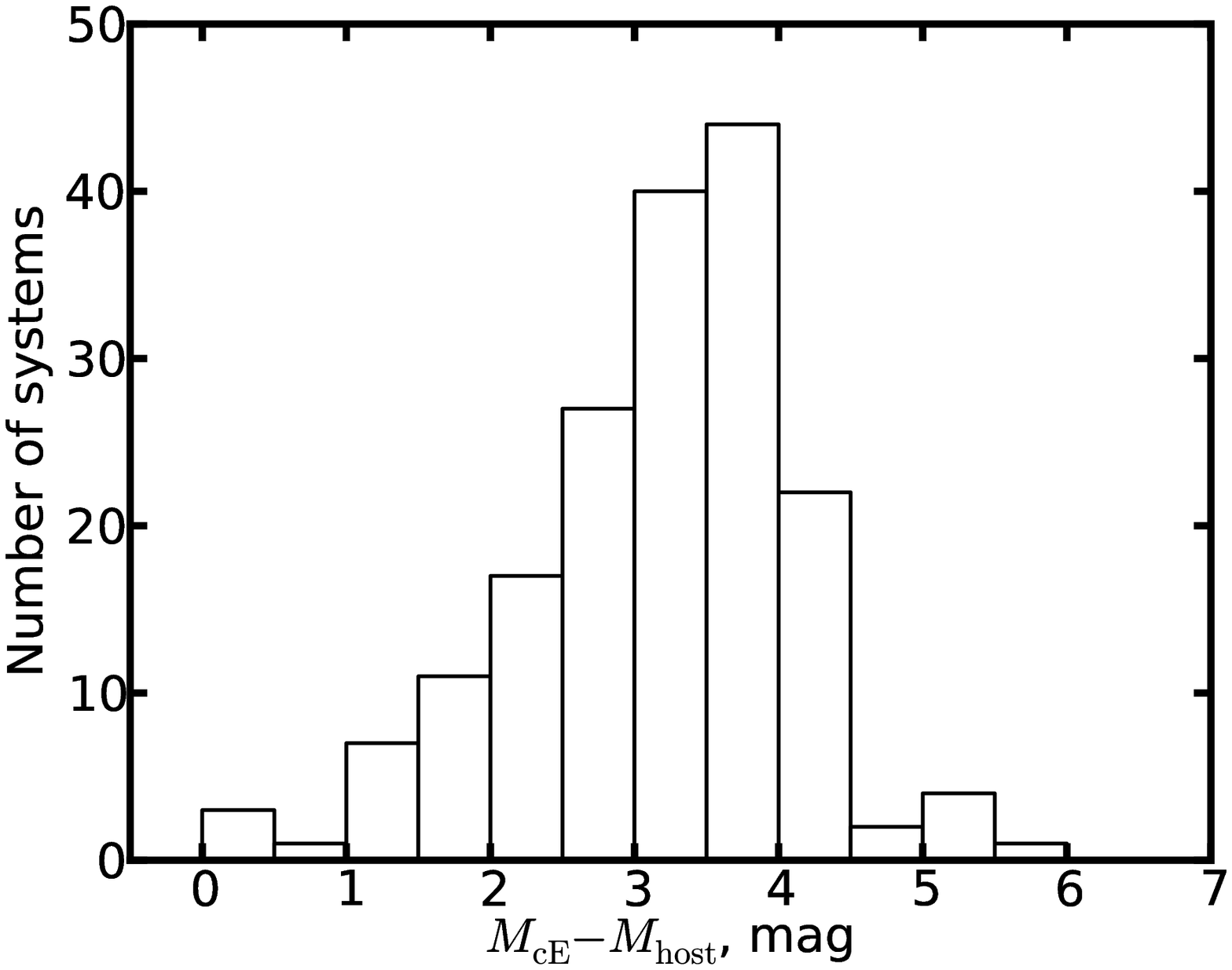}
\caption{\fontsize{8pt}{1.21em}\fontfamily{phv}\selectfont A distribution of $r$-band magnitude difference of cEs and their
identified host galaxies $M_{\rm{cE}}-M_{\rm{host}}$ reflecting the stellar
mass ratios. 
\label{figMMhost}}
\end{figure}
    \end{minipage}\hfill %
    \begin{minipage}{.45\textwidth}
\begin{figure}[H]
        \centering
\includegraphics[width=\hsize]{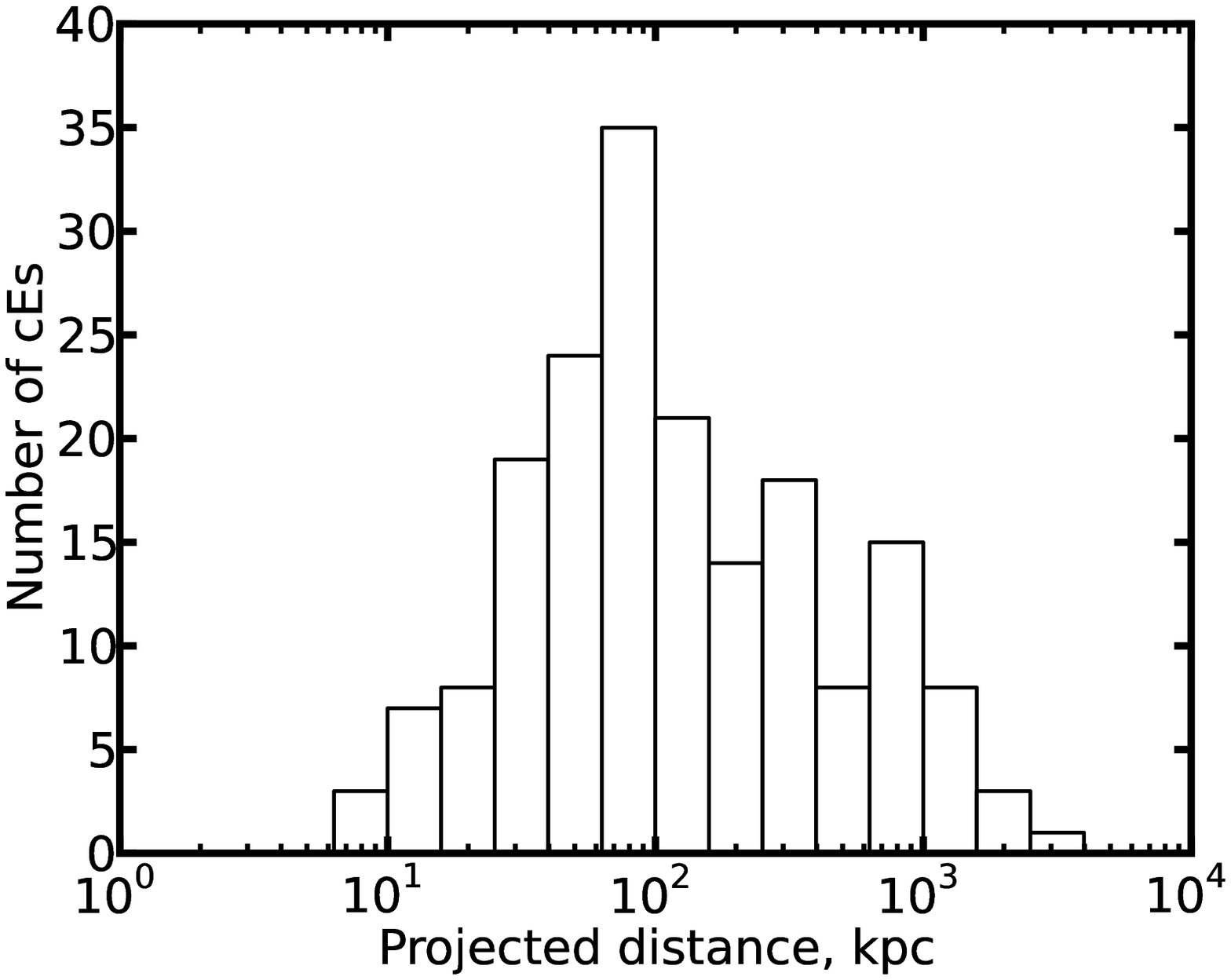}
\caption{\fontsize{8pt}{1.21em}\fontfamily{phv}\selectfont A distribution of projected distances between cEs and their
identified host galaxies. Isolated cE galaxies are excluded from the plot.
\label{figdproj}}
\end{figure}
    \end{minipage}

\begin{table}
\begin{tabular}{lcccccc}
\hline
ce\_objid & ra & dec & plate & mjd & fiberid & $\dots$ \\
\hline
587722953304375642 & 237.29089 & 0.21562 & 342 & 51691 & 548
\end{tabular}
\caption{\fontsize{8pt}{1.21em}\fontfamily{phv}\selectfont Properties of 195 cEs and their host galaxies found in this
study. Missing value in the {\fontseries{b}\selectfont host\_objid} column indicates that this row
corresponds to an isolated cE galaxy without an identified host. 
The complete table along with column descriptions is available in Microsoft
Excel format as a separate supplementary file.}
\end{table}

\vskip 15pt
\begin{multicols}{2}
\bibliographysm{cE_GALEXSDSS}
\bibliographystylesm{Science}
\end{multicols}

\end{document}